\documentclass[a4paper,11pt]{article}
\pdfoutput=1 

\usepackage{jinstpub} 
\usepackage{xspace}
\usepackage{orcidlink}

\usepackage{natbib}
\usepackage{subcaption}
\usepackage{lineno}

\newcommand{\htp}{\ensuremath{\mathrm{H}_2^+}\xspace}

\title{Computational Models for High-Power Cyclotrons and FFAs}

\author[a]{Andreas Adelmann  \orcidlink{0000-0002-7230-7007},\note{Corresponding author.}}
\author[b]{Chris T. Rogers,}

\affiliation[a]{Paul Scherrer Institut,\\Forschungsstrasse 111 CH-5232 Villigen, Switzerland}
\affiliation[b]{STFC Rutherford Appleton Laboratory,\\Harwell Science and Innovation Campus, Didcot, OX11 0QX, United Kingdom}
\emailAdd{andreas.adelmann@psi.ch, chris.rogers@stfc.ac.uk}

\abstract{A summary of numerical modeling capabilities regarding high power cyclotrons and fixed field alternating gradient machines is presented. This paper focuses on techniques made available by the OPAL simulation code.}


\begin{document}
\maketitle
\flushbottom

\section{Overview on Computational Models}
\label{sec:compmodels}
In all high-power particle accelerators "one of the major limitations is particle losses. Losses may be controlled, resulting in beam particles impinging on dedicated equipment such as collimators, or uncontrolled, resulting in beam particles striking other equipment around the accelerator. Uncontrolled losses can damage and activate any equipment in the accelerator and so must be minimized. Controlled losses need to be carefully considered and also minimized. The amount and cause of loss are investigated by modeling accelerators using simulation codes that model numerically the behaviour of beams. A review of available numerical codes can be found in the article of Smirnov~\cite{PhysRevAccelBeams.20.124801}. In this paper modeling capabilities available in OPAL are discussed in more detail \cite{OPAL-Manual}.

\subsection{Single particle modeling} 
For conventional cyclotrons (and FFAs) the single particle tool box is established 
and many different 
codes variants exists ~\cite{PhysRevAccelBeams.20.124801}. For cyclotrons and (horizontal FFAs) the existing 
tools seem to be comfortable and accurate. New machines like vertical FFAs,
currently studied for example at the Rutherford Appleton
Laboratory (RAL)~\cite{PhysRevAccelBeams.24.021601}, require non--trivial modifications
to the existing codes. These modifications are on the way for example in the code 
OPAL \cite{OPAL-Manual} and expected to be available in second quarter of 2022.

Recently, in the context of very high field and ultra
compact H$^-$ cyclotrons beam stripping losses of 
ion beams by interactions with residual gas and electromagnetic 
fields are evaluated~\cite{PhysRevAccelBeams.24.090101}. The beam stripping algorithm, implemented in OPAL, evaluates the interaction of hydrogen 
ions with residual gas and electromagnetic fields. In the first case, the cross 
sections of the processes are estimated according to the energy by means of analytical 
functions (see Sec. II-A c\cite{PhysRevAccelBeams.24.090101}). The implementation allows 
the user to set the pressure, temperature, and composition of the residual gas, which 
could be selected for the calculations as either molecular hydrogen (\htp) or dry air 
in the usual proportion. For precise simulations, a two-dimensional pressure field map 
from an external file can be imported into OPAL, providing more realistic vacuum 
conditions.

Concerning electromagnetic stripping, the electric dissociation lifetime is evaluated through
the theoretical formalism (see Sec. II-B \cite{PhysRevAccelBeams.24.090101}). 
In both instances, the individual probability at each integration step for every 
particle is assessed.

A stochastic process is used to evaluate if an interaction occurs. In this case the particle will be stripped and removed from the beam, or optionally transformed to a secondary heavy particle, dependent on the interaction. In this case, the secondary particle will continue its movement but with the new particle properties. 


\subsection{Large Scale Multiparticle Modeling} 
In general, modeling losses in high intensity accelerators require 3D space-charge and sufficient simulation particles. Recent investigations  \cite{MURALIKRISHNAN2021100094} propose a sparse grid-based adaptive noise reduction strategy for electrostatic particle-in-cell (PIC) simulations. By projecting the charge density onto sparse grids, high-frequency particle noise is reduced and hence an optimal number of grid points and simulation particles can be obtained. 
For a 3D Penning trap simulation,  a maximum speedup of 2.8 and 15 times memory reduction has been obtained. This method is already integrated into OPAL. 

\subsection{Surrogate Model Construction}
Cheap to evaluate surrogate models have gained a lot of interest lately. 
Statistical \cite{adelmann_2019} or machine learning techniques are used
\cite{info12090351}. These models can for example replace a computationally
heavy model in a multi-objective optimization \cite{adelmann-2020-1} 
or in the future be part of an on-line model. Some surrogate modeling algorithms may include an intrinsic estimator for the model uncertainty \cite{frey_2021}.

\section{Physics Modeling} 
In this section we show latest additions to the open source code OPAL \cite{OPAL-Manual} regarding cyclotron and FFA modeling capabilities. 

\subsection{Modeling H- Injection and Painting in Vertical and Horizontal FFAs}
Fixed Field Accelerators (FFAs) have fixed magnetic fields, like cyclotrons, but increase bending field with momentum and hence more compact designs can be realized. FFAs offer the power efficiency of cyclotrons combined with the energy reach of synchrotrons.

FFAs have never been used for high power proton acceleration, however in OPAL the necessary models are available for design. Single particle tracking has been benchmarked against the KURNS FFA \cite{Sheehy:2015cji}. A design for a 3-12 MeV H- FFA prototype ring is being pursued at RAL as a prototype for a MW-class neutron spallation source \cite{PhysRevAccelBeams.24.021601}. Scaling horizontal orbit excursion (hFFA) and a vertical orbit excursion (vFFA) FFA are both under consideration. Both are non--isochronous machines using RF cavities with variable resonant frequency. Injection is planned using charge exchange of H$^-$ to H$^+$ and phase space painting. 

In hFFAs, magnetic rigidity varies with radius. The dipole field varies as \cite{PhysRev.103.1837}
\begin{equation}
B_z(z=0) = B_0(\psi) \left(\frac{r}{r_0}\right)^k.
\end{equation}
$B_0(\psi)$ is the dipole field as a function of a normalised azimuthal coordinate $\psi$, $r$ is the radial coordinate, $r_0$ is a nominal (user-defined) radius, and $k$ is the field index. The field away from the midplane, at $z \neq 0$, may be calculated using a recursion relation arising from consideration of Maxwell's equations in free space. OPAL has capability to calculate the expansion to arbitrary order, within machine precision. The normalised azimuthal coordinate
\begin{equation}
\psi = \phi - \tan(\delta) \ln\left(\frac{r}{r_0}\right)
\end{equation}
is a measure of distance around the ring. Here $\phi$ is the geometrical azimuthal angle and $\delta$ is the spiral angle; for a sector FFA magnet $\delta = 0$ and $\psi = \phi$. The arrangement of fields in this way guarantees that single particle trajectories and optical parameters at all orders scale exactly with momentum.

In vFFAs, magnetic rigidity varies with height. As particles are accelerated, the closed orbit changes height. Successive acceleration kicks add incoherently, so overall the beam follows the closed orbit with no appreciable emittance growth. Rectangular vFFA magnets have been implemented in OPAL, with a dipole field that varies as \cite{PhysRevSTAB.16.084001}
\begin{equation}
B_0(x_{v}=0) = B_0(s_{v}) e^{mz_{v}}.
\end{equation}
$z_{v}$ is the height, $s_{v}$ is a nominal longitudinal coordinate and $x_{v}$ is a nominal horizontal coordinate in the rectangular coordinate system of the magnet. $B_0$ describes the dipole field variation with longitudinal distance. A $\tanh$ model is available for vFFA fields. $m$ is the vFFA field index, roughly equivalent to the field index $k$ in hFFAs. Fields away from the plane having $x_v=0$ are calculated using a field expansion derived from consideration of Maxwell's laws. It is noted that the focusing in the magnet body is, to linear order, skew quadrupole. The fringe field has solenoid components parallel to $s_v$ that may be significant for short magnets. This arrangement of fields guarantees that trajectories and optical functions are identical as momentum increases, barring a vertical displacement. In particular, the path length of the beam is independent of momentum, the momentum compaction factor is exactly 0 and ultra-relativistic particles are isochronous.

In order to model injection into the FFA, OPAL was extended with models for:
\begin{itemize}
\item horizontal \& vertical FFA magnets as described above;
\item variable frequency RF cavities;
\item arbitrary order multipoles with maxwellian fringe fields;
\item foil model (scattering and energy loss);
\item pulsed injected beam; and
\item pulsed multipoles.
\end{itemize}
All but the latter two features are available in the latest version of OPAL. This enabled a fully four-dimensional simulation of the injection system, including consideration of effects such as appropriate phasing of the pulsed dipoles and transverse breathing of the beam arising due to initial longitudinal mismatch at injection.

\begin{figure}
\begin{subfigure}{0.5\textwidth}
\vspace{-3.5cm}
\includegraphics[width=\textwidth, trim={5cm 0 4cm 0}]{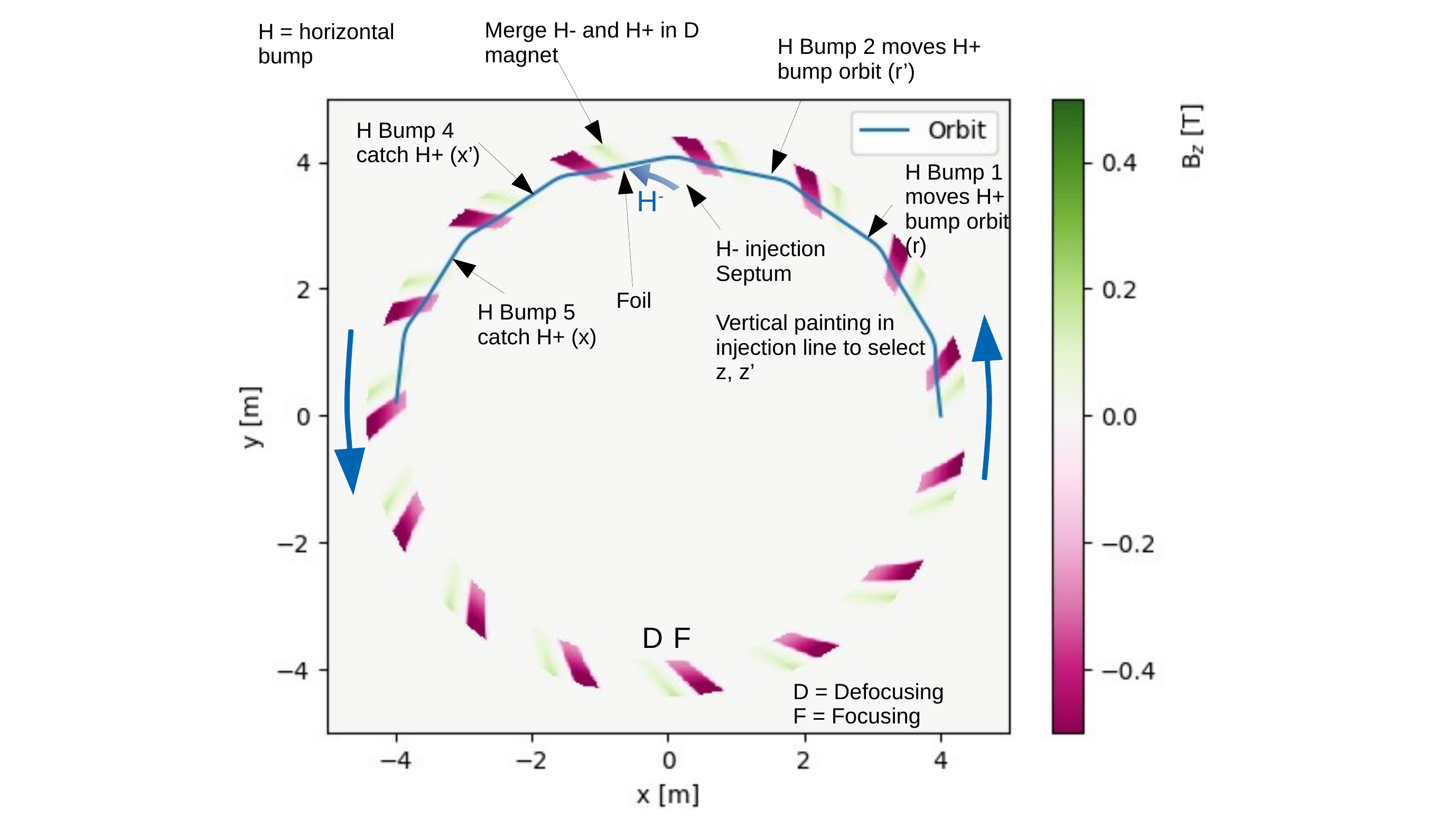}
\end{subfigure}
\begin{subfigure}{0.5\textwidth}
\includegraphics[width=0.94\textwidth]{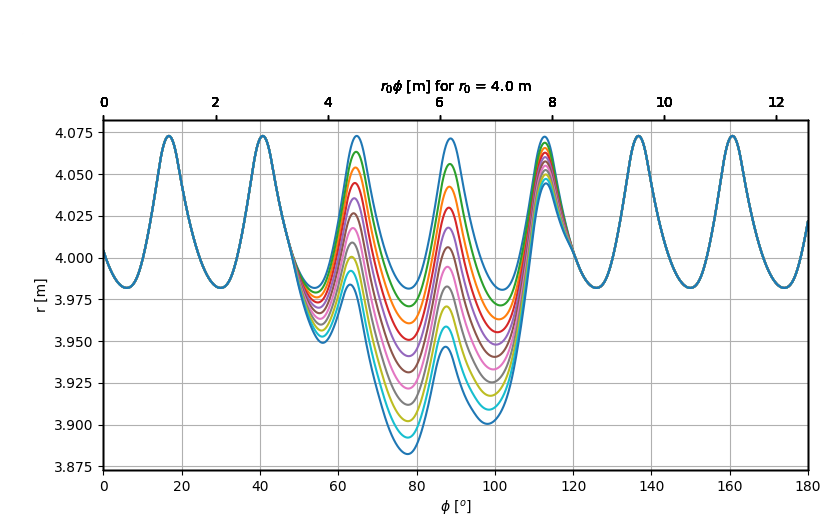}
\includegraphics[width=\textwidth]{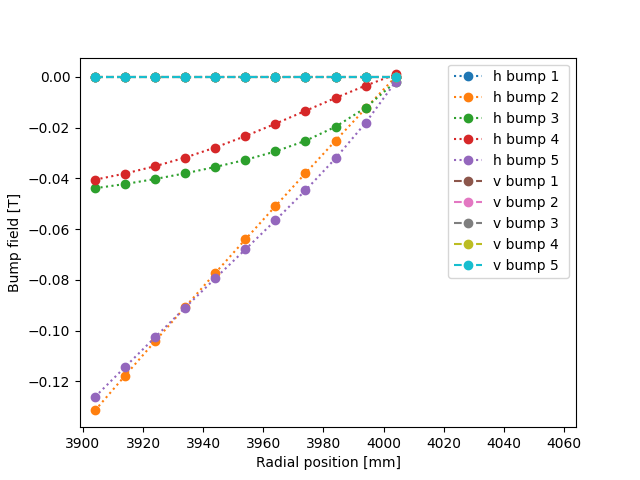}
\end{subfigure}
\caption{Injection system for the hFFA (Left) field map of the hFFA, calculated using OPAL, with labels indicating the position of injection equipment (top right) closed orbits for different bump magnets (bottom right) required bump magnet fields.}
\label{fig:hffainjection}
\end{figure}

\begin{figure}
\includegraphics[width=0.5\textwidth]{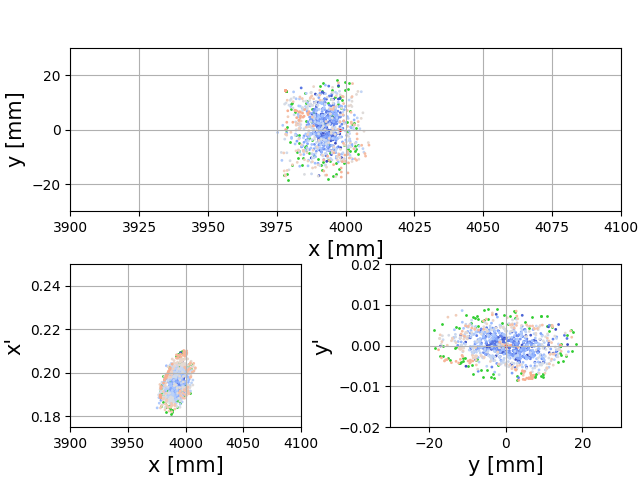}
\includegraphics[width=0.5\textwidth]{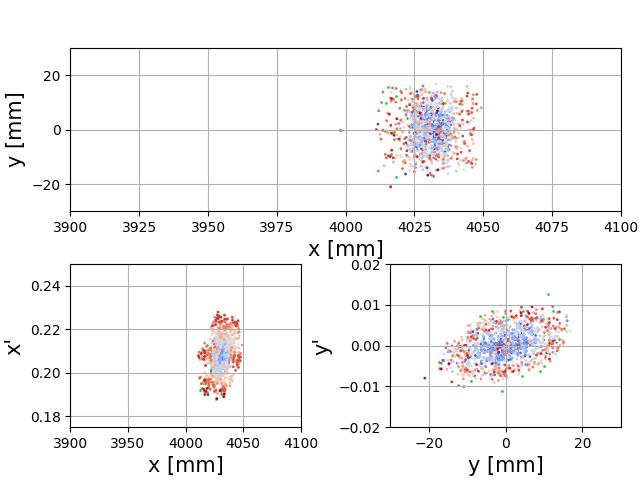}
\caption{Beam (left) after injection is completed, but still on a distorted orbit (right) following collapse of the bump. $x$ is the position of the beam relative to the ring centre and $y$ is the height of the particle above the midplane. Particles are coloured according to the injection turn.}
\label{fig:hffainjectedbeam}
\end{figure}

As an example, a schematic of an injection system and associated parameters for the 3-12 MeV test ring is shown for a horizontal FFA in Fig. \ref{fig:hffainjection}. Owing to the compact nature of the ring, the injection system is spread across a number of cells. H$^-$ are brought into the ring and onto a foil. Bump magnets in the ring distort the proton closed orbit so that particles passing through the foil are returned to a nominal closed orbit. The foil is placed inside the defocusing (D) dipole magnet so that the distorted H$^+$ closed orbit and H$^-$ beam, initially separated, are brought onto the same trajectory.  Electrons are stripped from the H$^-$ leaving H$^+$ (protons). The bump magnets are slowly varied, so that the proton closed orbit is moved away from the injection point for the H$^-$ and newly injected particles are at higher horizontal amplitude. In the H$^-$ injection line, pulsed magnets move the H$^-$ upwards so that newly injected particles are at higher vertical amplitude. Overall, a correlation is introduced between horizontal and vertical amplitude. Sample trajectories and bump magnet field strengths for the magnets in the ring are shown in Fig. \ref{fig:hffainjection}. In this example vertical bumpers are not considered - they are all kept at 0 T field. The beam following injection is shown in fig. \ref{fig:hffainjectedbeam}.

\subsection{Beam stripping interactions}
 Beam transmission optimization and loss characterization, where beam stripping interactions are a key issue, play an important role in the design and operation of compact cyclotrons. A beam stripping model has been implemented in the three-dimensional object-oriented parallel code OPAL-cycl, a flavor of the OPAL framework. The model includes Monte Carlo methods for interaction with residual gas and dissociation by electromagnetic stripping. The model has been verified with theoretical models and it has been applied to the AMIT cyclotron according to design conditions \cite{PhysRevAccelBeams.24.090101}.
 
\subsection{Spiral inflector modeling}
\begin{figure}
\begin{subfigure}{1.\textwidth}
\centering\includegraphics[width=0.6\textwidth]{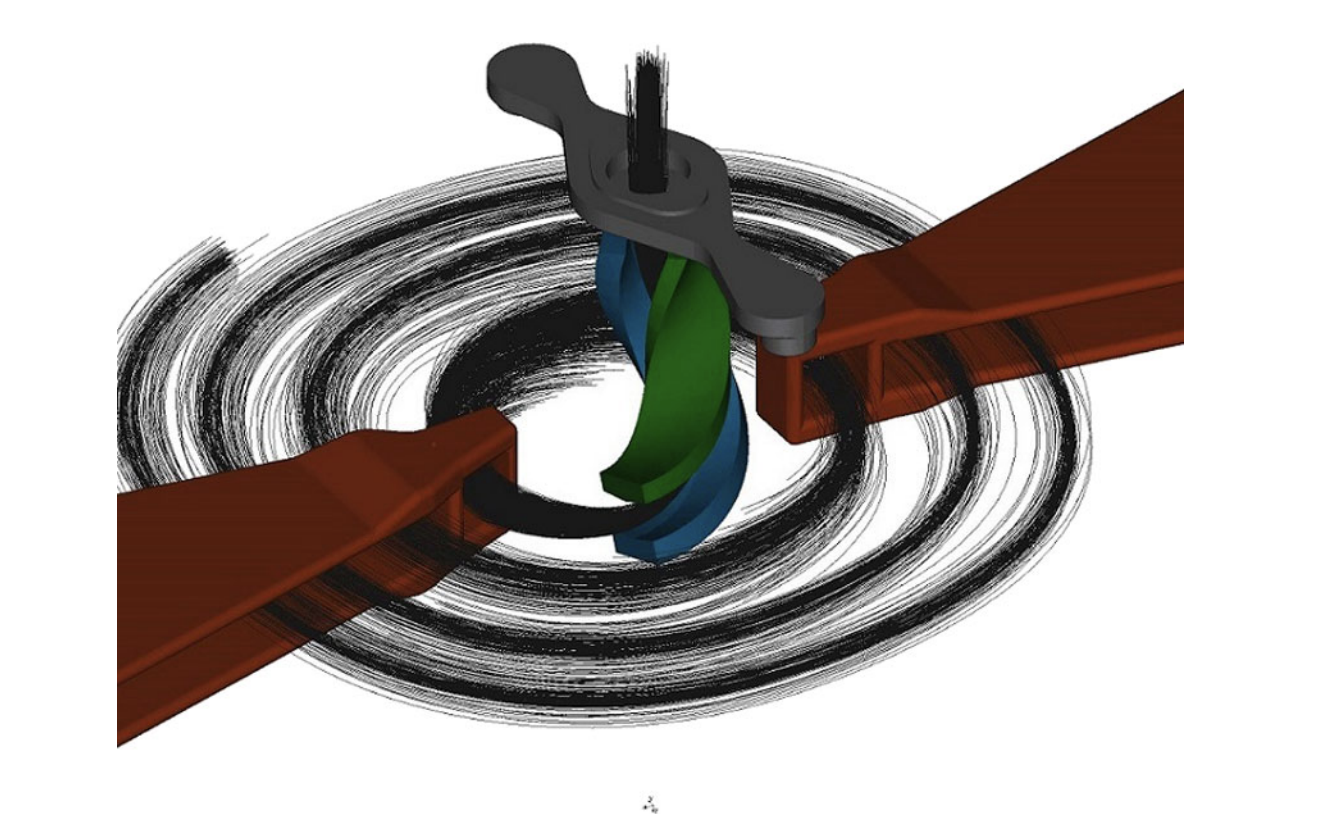}
\end{subfigure}
\caption{Spiral inflector with selected particle trajectories
from an OPAL simulation. The beam enters axially (from the top)
through an aperture (grey) and is bent into the mid-plane by a combination of the electrostatic field generated by the spiral
electrodes (green and blue) and the cyclotron's main magnetic field.
Then it is accelerated by the two Dees (copper, Dummy-Dees not shown) \cite{winklehner:spiral}.}
\label{fig:spiralinject}
\end{figure}

In \cite{winklehner:spiral} a spiral inflector model implemented in OPAL is presented, that enables us to run
highly realistic simulations of the spiral inflector system of a compact cyclotron (c.f.\ Fig.\ \ref{fig:spiralinject}). A new geometry class and field solver 
can handle the complicated boundary conditions posed by
the electrode system in the central region of the cyclotron both in terms of particle termination, and
calculation of self-fields. Results are benchmarked against the analytical solution of a coasting beam. As
a practical example, the spiral inflector and the first revolution in a $1$ MeV/amu test cyclotron, located at
Best Cyclotron Systems, Inc., are modeled and compared to the simulation results \cite{PhysRevAccelBeams.20.124201,Alonso_2015}.\ In conclusion, OPAL can
handle realistic and arbitrary boundary geometries. Simulated injection efficiencies and beam
shape compare well with measured efficiencies and a preliminary measurement of the beam distribution
after injection.

\subsection{Neighboring Turn Modeling}
This article presents a hardware architecture independent implementation of an adaptive mesh refinement Poisson solver that is integrated into the electrostatic Particle-In-Cell beam dynamics code OPAL. The Poisson solver is solely based on second generation Trilinos packages to ensure the desired hardware portability. Based on the massively parallel framework AMREX, formerly known as BoxLib, the new adaptive mesh refinement interface provides several refinement policies in order to enable precise large-scale neighbouring bunch simulations in high intensity cyclotrons. The solver is validated with a built-in multigrid solver of AMREX and a test problem with analytical solution. The parallel scalability is presented as well as an example of a neighbouring bunch simulation that covers the scale of the later anticipated physics simulation \cite{frey2021architecture}.

\begin{figure}
\begin{subfigure}{1.\textwidth}
\centering\includegraphics[width=0.6\textwidth]{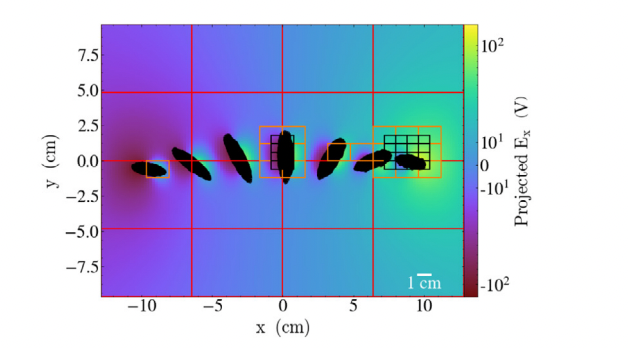}
\end{subfigure}
\caption{
Integrated projection of the electric field component $E_{x}$ onto the xy-plane showing 7 adjacent particle bunches \cite{frey2021architecture}.}
\label{fig:neigh-turns}
\end{figure}

\section{Path Forward}
\begin{figure}[htbp]
\centering 
\includegraphics[width=.8\textwidth,clip]{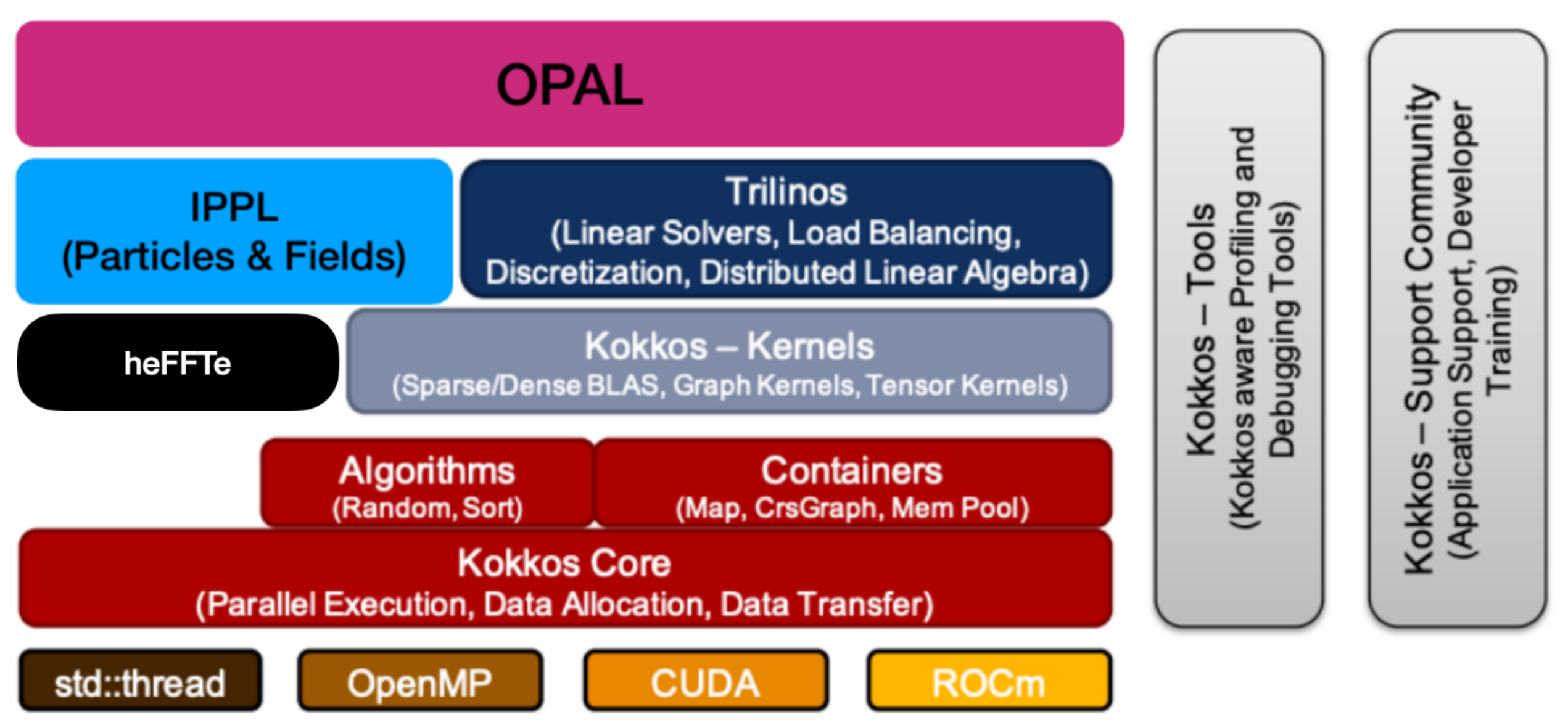}
\caption{\label{fig:opalx} Outlook of the future OPAL architecture, targeting in a performance portable way future exascale architectures.}
\end{figure}

While statistical and  machine learning techniques have a lot of potential, high fidelity physics simulations will always be used
to, for example, produce the training set. In case of high-intensity machines we will need large numbers of particles and the 
associated fine mesh to solve the PDE in question. It is imperative that we make use of existing and future high performance infrastructure. A performance portable implementation \cite{frey2021architecture} is of utmost importance. The OPAL collaboration \cite{OPAL-Manual} is in
the progress to completely rewrite the code according to the sketch in Fig. \ref{fig:opalx}. With this new architecture we will be able to make efficient use of Exascale-Architecture that will come online soon.
The core algorithms of OPAL are already performance portable as demonstrated in \cite{2205.11052}.


\acknowledgments
The authors acknowledge the OPAL developer team for their continued support of this 
open source, community-driven code.

\bibliography{bibliography}

\begin{thebibliography}{17}
\providecommand{\natexlab}[1]{#1}
\providecommand{\url}[1]{\texttt{#1}}
\expandafter\ifx\csname urlstyle\endcsname\relax
  \providecommand{\doi}[1]{doi: #1}\else
  \providecommand{\doi}{doi: \begingroup \urlstyle{rm}\Url}\fi

\bibitem[Smirnov(2017)]{PhysRevAccelBeams.20.124801}
V.~Smirnov.
\newblock Computer codes for beam dynamics analysis of cyclotronlike
  accelerators.
\newblock \emph{Phys. Rev. Accel. Beams}, 20:\penalty0 124801, 12 2017.
\newblock \doi{10.1103/PhysRevAccelBeams.20.124801}.
\newblock URL
  \url{https://link.aps.org/doi/10.1103/PhysRevAccelBeams.20.124801}.

\bibitem[OPA(2021)]{OPAL-Manual}
{The OPAL Framework: Version 2.4}, 2021.
\newblock \url{http://amas.web.psi.ch/opal/Documentation/2.4/index.html}.

\bibitem[Machida et~al.(2021)Machida, Kelliher, Lagrange, and
  Rogers]{PhysRevAccelBeams.24.021601}
S.~Machida, D.~J. Kelliher, J-B. Lagrange, and C.~T. Rogers.
\newblock Optics design of vertical excursion fixed-field alternating gradient
  accelerators.
\newblock \emph{Phys. Rev. Accel. Beams}, 24:\penalty0 021601, 2 2021.
\newblock \doi{10.1103/PhysRevAccelBeams.24.021601}.
\newblock URL
  \url{https://link.aps.org/doi/10.1103/PhysRevAccelBeams.24.021601}.

\bibitem[Calvo et~al.(2021)Calvo, Podadera, Gavela, Oliver, Adelmann,
  Snuverink, and Gsell]{PhysRevAccelBeams.24.090101}
P.~Calvo, I.~Podadera, D.~Gavela, C.~Oliver, A.~Adelmann, J.~Snuverink, and
  A.~Gsell.
\newblock Beam stripping interactions in compact cyclotrons.
\newblock \emph{Phys. Rev. Accel. Beams}, 24:\penalty0 090101, 11 2021.
\newblock \doi{10.1103/PhysRevAccelBeams.24.090101}.
\newblock URL
  \url{https://link.aps.org/doi/10.1103/PhysRevAccelBeams.24.090101}.

\bibitem[Muralikrishnan et~al.(2021)Muralikrishnan, Cerfon, Frey, Ricketson,
  and Adelmann]{MURALIKRISHNAN2021100094}
Sriramkrishnan Muralikrishnan, Antoine~J. Cerfon, Matthias Frey, Lee~F.
  Ricketson, and Andreas Adelmann.
\newblock Sparse grid-based adaptive noise reduction strategy for
  particle-in-cell schemes.
\newblock \emph{Journal of Computational Physics: X}, 11:\penalty0 100094,
  2021.
\newblock ISSN 2590-0552.
\newblock \doi{https://doi.org/10.1016/j.jcpx.2021.100094}.
\newblock URL
  \url{https://www.sciencedirect.com/science/article/pii/S2590055221000111}.

\bibitem[Adelmann(2019)]{adelmann_2019}
Andreas Adelmann.
\newblock On nonintrusive uncertainty quantification and surrogate model
  construction in particle accelerator modeling.
\newblock \emph{SIAM/ASA Journal on Uncertainty Quantification}, 7\penalty0
  (2):\penalty0 383--416, 2019.

\bibitem[Bellotti et~al.(2021)Bellotti, Boiger, and Adelmann]{info12090351}
Renato Bellotti, Romana Boiger, and Andreas Adelmann.
\newblock Fast, efficient and flexible particle accelerator optimisation using
  densely connected and invertible neural networks.
\newblock \emph{Information}, 12\penalty0 (9), 2021.
\newblock \doi{10.3390/info12090351}.
\newblock URL \url{https://www.mdpi.com/2078-2489/12/9/351}.

\bibitem[Edelen et~al.(2020)Edelen, Neveu, Huber, Frey, and
  Adelmannn]{adelmann-2020-1}
Auralee Edelen, Nicole Neveu, Yannick Huber, Matthias Frey, and Andreas
  Adelmannn.
\newblock Machine learning to enable orders of magnitude speedup in
  multi-objective optimization of particle accelerator systems'.
\newblock \emph{Phys. Rev. AB}, 23:\penalty0 044601, 2020.
\newblock \doi{10.1103/PhysRevAccelBeams.23.044601}.
\newblock URL
  \url{https://link.aps.org/doi/10.1103/PhysRevAccelBeams.23.044601}.

\bibitem[Frey and Adelmann(2021)]{frey_2021}
Matthias Frey and Andreas Adelmann.
\newblock Global sensitivity analysis on numerical solver parameters of
  particle-in-cell models in particle accelerator systems.
\newblock \emph{Computer Physics Communications}, 258:\penalty0 107577, 2021.
\newblock ISSN 0010-4655.
\newblock \doi{https://doi.org/10.1016/j.cpc.2020.107577}.
\newblock URL
  \url{http://www.sciencedirect.com/science/article/pii/S0010465520302770}.

\bibitem[Sheehy et~al.(2015)]{Sheehy:2015cji}
Suzanne Sheehy et~al.
\newblock {Progress on Simulation of Fixed Field Alternating Gradient
  Accelerators}.
\newblock In \emph{{6th International Particle Accelerator Conference}}, page
  MOPJE077, 2015.
\newblock \doi{10.18429/JACoW-IPAC2015-MOPJE077}.

\bibitem[Symon et~al.(1956)Symon, Kerst, Jones, Laslett, and
  Terwilliger]{PhysRev.103.1837}
K.~R. Symon, D.~W. Kerst, L.~W. Jones, L.~J. Laslett, and K.~M. Terwilliger.
\newblock Fixed-field alternating-gradient particle accelerators.
\newblock \emph{Phys. Rev.}, 103:\penalty0 1837--1859, Sep 1956.
\newblock \doi{10.1103/PhysRev.103.1837}.
\newblock URL \url{https://link.aps.org/doi/10.1103/PhysRev.103.1837}.

\bibitem[Brooks(2013)]{PhysRevSTAB.16.084001}
Stephen Brooks.
\newblock Vertical orbit excursion fixed field alternating gradient
  accelerators.
\newblock \emph{Phys. Rev. ST Accel. Beams}, 16:\penalty0 084001, Aug 2013.
\newblock \doi{10.1103/PhysRevSTAB.16.084001}.
\newblock URL \url{https://link.aps.org/doi/10.1103/PhysRevSTAB.16.084001}.

\bibitem[Winklehner et~al.(2017{\natexlab{a}})Winklehner, Adelmann, Gsell,
  Kaman, and Campo]{winklehner:spiral}
Daniel Winklehner, Andreas Adelmann, Achim Gsell, Tulin Kaman, and Daniela
  Campo.
\newblock Realistic simulations of a cyclotron spiral inflector within a
  particle-in-cell framework.
\newblock \emph{Physical Review Accelerators and Beams}, 20\penalty0
  (12):\penalty0 124201, 12 2017{\natexlab{a}}.
\newblock \doi{10.1103/PhysRevAccelBeams.20.124201}.
\newblock URL
  \url{https://link.aps.org/doi/10.1103/PhysRevAccelBeams.20.124201}.

\bibitem[Winklehner et~al.(2017{\natexlab{b}})Winklehner, Adelmann, Gsell,
  Kaman, and Campo]{PhysRevAccelBeams.20.124201}
Daniel Winklehner, Andreas Adelmann, Achim Gsell, Tulin Kaman, and Daniela
  Campo.
\newblock Realistic simulations of a cyclotron spiral inflector within a
  particle-in-cell framework.
\newblock \emph{Phys. Rev. Accel. Beams}, 20:\penalty0 124201, Dec
  2017{\natexlab{b}}.
\newblock \doi{10.1103/PhysRevAccelBeams.20.124201}.
\newblock URL
  \url{https://link.aps.org/doi/10.1103/PhysRevAccelBeams.20.124201}.

\bibitem[Alonso et~al.(2015)Alonso, Axani, Calabretta, Campo, Celona, Conrad,
  Day, Castro, Labrecque, and Winklehner]{Alonso_2015}
J.~Alonso, S.~Axani, L.~Calabretta, D.~Campo, L.~Celona, J.M. Conrad, A.~Day,
  G.~Castro, F.~Labrecque, and D.~Winklehner.
\newblock The isodar high intensity h2+ transport and injection tests.
\newblock \emph{Journal of Instrumentation}, 10\penalty0 (10):\penalty0 T10003,
  oct 2015.
\newblock \doi{10.1088/1748-0221/10/10/T10003}.
\newblock URL \url{https://dx.doi.org/10.1088/1748-0221/10/10/T10003}.

\bibitem[Frey et~al.(2021)Frey, Adelmann, and Locans]{frey2021architecture}
Matthias Frey, Andreas Adelmann, and Uldis Locans.
\newblock On architecture and performance of adaptive mesh refinement in an
  electrostatics particle-in-cell code (vol 247, 106912, 2020).
\newblock \emph{COMPUTER PHYSICS COMMUNICATIONS}, 265, 2021.

\bibitem[Muralikrishnan et~al.(2022)Muralikrishnan, Frey, Vinciguerra,
  Ligotino, Cerfon, Stoyanov, Gayatri, and Adelmann]{2205.11052}
Sriramkrishnan Muralikrishnan, Matthias Frey, Alessandro Vinciguerra, Michael
  Ligotino, Antoine~J. Cerfon, Miroslav Stoyanov, Rahulkumar Gayatri, and
  Andreas Adelmann.
\newblock Alpine: A set of performance portable plasma physics particle-in-cell
  mini-apps for exascale computing, 2022.
\newblock URL \url{arXiv:2205.11052}.

\end{thebibliography}

%
%
%
%
%
%
%
%
\end{document}